\documentclass[manuscript]{aastex631}

\usepackage{graphicx}	
\usepackage{amsmath}	
\usepackage{amssymb}	
 \usepackage{color}
\shorttitle{Stability and Planet-Formation in Polar Circumbinary Disc}
\shortauthors{Ying Wang et al.}
\graphicspath{{./}{figures/}}

\begin{document}

\title{Dynamical Stability of Polar Circumbinary Orbits and Planet-Formation in Planetary Disc of 99 Herculis}

\correspondingauthor{Wei Sun,Fu-yao Liu}
\email{sunweiay@163.com, liufuyao2017@163.com,wangying424524@163.com,zhoujil@nju.edu.cn,ming.yang@nju.edu.cn}

\author[0000-0003-0506-054X]{Ying Wang}
\affiliation{School of Mathematics, Physics and Statistics, Shanghai University of Engineering Science, Shanghai 201620, China}

\author[0000-0003-2620-6835]{Wei Sun}
\affiliation{School of Mathematics, Physics and Statistics, Shanghai University of Engineering Science, Shanghai 201620, China}

\author{Ji-lin Zhou}
\affiliation{School of Astronomy and Space Science and Key Laboratory of Modern Astronomy and Astrophysics in Ministry of Education, Nanjing University, Nanjing 210093, China}

\author[0000-0002-6926-2872]{Ming Yang}
\affiliation{College of Surveying and Geo-Informatics, Tongji University, Shanghai 200092, China}

\author{Fu-yao Liu}
\affiliation{School of Mathematics, Physics and Statistics, Shanghai University of Engineering Science, Shanghai 201620, China}



\begin{abstract}

A possible polar-ring debris disc, the dynamics of which can be described by the outer hierarchical restricted three-body problem, has been detected in 99 Herculis. An empirical formula on the minimum radius beyond which test particles in polar orbits can keep stable within ${10^7}$ binary periods is provided through the numerical fitting, applying to the binary eccentricity $e_{1}  \in \left[ {0,0.8} \right)$ and the mass ratio of binary $ \lambda \in \left[ {0.1,1} \right]$, where $ \lambda = m_0/m_1$ (${m_0}$ and ${m_1}$ represent the masses of the two binary stars). The polar planetary disc has the lowerest statistical accretion efficiency and moderate impact frequency of collisions among planetesimals (with a radius of 1-10km) compared to that in the circumbinary coplanar disc and the standard disc around the single host star. Colliding timescales in the circumbinary disk (both polar and coplanar configuration) are longer than $10^7$ yr exceeding the dissipation timescales of the gas disc. The stochastic simulations show that successive collisions cannot make planetesimal grow up which may explain the formation of the debris disc observed in 99 Herculis.

\end{abstract}



\section{Introduction} \label{sec:intro}
Up until January 2023, approximately 737 planets have been discovered in binary systems\footnote{http://exoplanets.org/}. These planets in binary systems can be classified into three types according to their orbital configuration, S-type (satellite-type), P-type (planet-type) and L-type (libration-type) \citep{dvorak1986critical}. Planets in S-type orbits, which are also called circumprimary planets, encircle one of the stellar binary components with the second star considered to be a perturber. Planets in P-type orbits, also known as circumbinary planets, encircle both members of a binary. The L-type planets librate around the Lagrangian equilibrium points $L_{4}$ or $L_{5}$, which can stably exist if the binary mass satisfies $\mu < 0.04$, where $\mu= {m_{0}}/({m_{0}+m_{1}})$ . The currently observed exoplanets in circumbinary systems include approximately 75 P-type planets, others are S-type planets. \footnote{http://exoplanetarchive.ipac.caltech.edu/index.html}

Planetary formation in binary systems is different from that in single-star systems because of the disturbance from the movements of binary stars. The planetary disc will be truncated during several binary periods \citep{artymowicz1994dynamics}, and forced into eccentric and processional motions \citep{larwood1996tidally,paardekooper2008planetesimal,fragner2010evolution}. The planets can also be forced into noncircular orbits by the disc eccentricity. The inward migration of planets will be stalled upon entrance to the tidally-truncated inner cavity in hydrodynamical simulations on the formation of P-type planets in Kepler 16, 34 and 35 systems \citep{pierens2013migration,pelupessy2013formation}.  2D locally isothermal hydrodynamical simulations of circumbinary discs with embedded planets were performed by \citet{penzlin2021parking}. The results strongly support the assumption that planets migrate to their present locations due to planet-disc interaction. During the migration, circumbinary planets could be captured into mean motion resonances, which may be associated with the final locations \citep{gianuzzi2023circumbinary}. 

Some P-type planets were detected close to the edges of stable regions, where the disturbance is remarkably powerful to form planets, even considering the most favorable case of $100\%$ efficient dust accretion \citep{moriwaki2004planetesimal,paardekooper2012not,meschiari2012circumbinary,martin2014planets}. Moreover, the self-gravity of the circumbinary disc can excite the eccentricities and prevent a full alignment of the planetesimal pericenters, thus resulting in sufficiently large impact velocities among planetesimals that damage the impacting planetesimals in the current location of circumbinary planets \citep{marzari2013influence}. The accretion of planetesimals is possible for $a_{2}/a_{1}>20$ in Kepler 16 \citep{paardekooper2012not,marzari2012lunar,meschiari2012circumbinary}, where $a_{1}$ and $a_{2}$ are the semi-major axes of binary and planetesimals respectively. The gravity of an axisymmetric disc strongly suppresses the eccentricities of planetesimals beyond $a_{2}/a_{1} \approx$ 10-20, facilitating the easy growth of 1-$10^{2}$ km objects \citep{rafikov2013building}. The critical radial distance beyond which planetesimal accretion is possible increases with rising binary eccentricities and decreasing mass ratio based on the examination of the relative velocities among accreting planetesimals \citep{scholl2007relative}. 

In addition to the difficulties of circumbinary planets' formation at observed locations,  the stable region around the binary where planets can survive for long times must be identified as a fundamental question of celestial mechanics. \citet{holman1999long} simulated the empirical criteria for the largest and the smallest stable orbits of test particles in the orbital planes of S-type and P-type binaries within $10^{4} T_1$ ($T_{1}$ is the binary period) in the range $0.0 \le e_{1} \le$ 0.7-0.8 and $0.1 \le \mu \le 0.5$. Many factors could influence the stability of planets during and after the post-oligarchic evolution. 
\citet{hong2019stability} extended the analysis of the chaotic region of coplanar P-type orbits by \citet{dvorak1986critical} to the un-restricted three-body problem and counter-rotating orbits. Mean motion resonances between P-type planets can interact with the binary via resonant and secular effects creating additional instabilities and driving chaos in multi-planet resonant systems \citep{sutherland2019instabilities}. \citet{thun2018migration} found that massive planets can significantly alter the disc structure and remain on nearly circular orbits based on the planet-disc mass ratio, while low-mass planets are strongly influenced by the disc, with eccentricities excited to high values. If additional planets formed in the circumbinary disc, planet-planet scattering which takes place near the location of the currently discovered circumbinary planets, left a single planet with low eccentricity with $90\%$ possibility \citep{gong2016p}.

Most of the circumbinary planets were detected by transiting and eclipse time variation. Detected circumbinary planets were usually in the coplanar plane with binary orbits due to the restrictions of the two observation methods. Interestingly, several misaligned circumbinary planetary discs have been detected. The precessional circumbinary-ring model which is mildly misaligned with the binary orbital plane by $10^\circ \sim 20^\circ$ has successfully interpreted the observations of KH 15D from 1995 to 2012 \citep{chiang2004circumbinary,winn2004kh,capelo2012locating}. 

\citet{lacour2016m} found that the binary orbital plane of HD 142527 inclines $70^\circ$ relative to the outer circumbinary disc, which was considered as a transition disc. Through 3D hydrodynamical simulations on HD142527, \citet{price2018circumbinary} confirmed that all of the main observational features such as the spirals, shadows, and horseshoe can be explained by the interaction between the disc and the inner binary. However, there is no consensus on the inclination of the disc which was considered to account for the optical asymmetry of dust. Different methods find different solutions, which mainly range from $20^ \circ $ to $28^ \circ $\citep{avenhaus2017exploring, hunziker2021hd}.

The circumbinary debris disc in the 99 Herculis system was resolved, which may move in the plane perpendicular to the binary pericenter direction \citep{kennedy201299}. The misalignment $\delta i$ between the young circumbinary protoplanetary disc around GG Tau A binary system in the quadruple system GG Tau and the binary orbit is approximately $25^\circ \sim 50^\circ$ \citep{cazzoletti2017testing,aly2018secular}.
Each stellar component of IRS43 has its circumstellar disc, and both are surrounded by a highly-inclined circumbinary disc ($>60^\circ$) \citep{brinch2016misaligned}. An unusual gas-rich circumbinary disc in the young HD 98800 system is probably in the polar configuration based on the simulation of the disc dynamics, and the physical properties of such disc are similar to those around young single stars \citep{verrier2008hd,Kennedy_2019}. The disc size (5-5.5 au) make it one of the smallest discs known\citep{ribas2018long}. \citet{ziglin1975secular} described the doubly averaged outer restricted elliptic three-body problem considering quadrupole approximation for the first time. HD 98800, which is a quadruple system, can be well approximated as a hierarchical triple-star system. High-inclination particles were found in long-term stable orbits inclined by $55^\circ\sim 135^\circ$ to the inner binary. During the study on disc formation, \citet{verrier2009high} found that inclination variations and nodal precession caused by the inner binary for certain initial longitudes can suppress Kozai cycles that would otherwise occur due to the outer star in the hierarchical three-body problem. \citet{farago2010high} provided a complete analytical description of the test particle in the secular and quadrupolar approximations of the outer hierarchical three-body problem using a vectorial formalism.

Given the stability of the inclined orbits, \citet{pilat2003stability} recorded the escape time for inclined P-type orbits ($ 0^\circ \le i \le 50^\circ$) in equal-mass binary systems with the binary eccentricity ranging from 0 to 0.5 within the integration time ${5\times10^4}{T_1}$, and distinguished different types of motions using the fast Lyapunov Indicator. \citet{doolin2011dynamics} studied the stability of inclined orbits of circumbinary test particles and found the critical radius of stable planets in polar orbits is smaller than that in coplanar orbits. Considering the mass of the planet as well as the interaction between the planet and binary, \citet{Chen2019MNRAS} extend the polar orbit to the generalized polar orbit by stationary inclinations, where the precession rates of the binary and planet are the same, and the relative inclination between the orbital plane of the planet and the binary are fixed. In contrast to that retrograde circulating orbits are usually the most stable around binary with small eccentricity, polar planets around highly eccentric are the most stable\citep{Chen2020MNRAS}.
N-body simulations were conducted to scan $\Delta e$, $\Delta \Omega_2$, and chaos indicators to study the global stability for different $\lambda $ of the binary and initial $\Omega$ of the polar orbits by \citet{cuello2019planet}.

However, an empirical formula for the critical radius of stable polar orbits has not been obtained, and the possibility of planet formation in circumbinary polar discs remains unknown. The paper aims to address the two issues. Firstly, the motions of test particles in polar orbits were briefly described considering the analysis of the elliptically restricted three-body problem in Section \ref{sec:dynamics}, which displays a libration mechanism in the longitude of the ascending node and the inclination relative to the plane of the binary. Then, an empirical formula of the stable boundary of circumbinary test particles in polar orbits with the longest integration time ${10^7}{T_1}$ was presented in Section \ref{sec:timescale}. The binary eccentricity is in the range of $\left[ {0,0.8} \right)$, and the mass ratio of binary star $\lambda = \left[ {0.1,1} \right]$. Lastly, statistical analyses and stochastic simulation of collisions among planetesimals in polar-ring discs were conducted in comparison with the results of the coplanar circumbinary disc and standard disc around the single star in Section \ref{sec:formation}.

\section{Circumbinary polar motion in the outer-restricted hierarchical three-body problem}
\label{sec:dynamics}
The complete Hamiltonian of the hierarchical three-body system can be described in Jacobian coordinates. The hierarchical three-body secular approximation \citep{harrington1968dynamical,harrington1969stellar} can be obtained by adopting Delaunay's canonical elements, considering the quadrupole moment, and averaging Hamiltonian over the short timescales by von Ziepel transformation \citep{kozai1962secular,harrington1968dynamical}. In the outer-restricted three-body problem, the outer body is assumed to be a massless test particle revolving around the binary pair, thus the outer body does not affect the inner orbit. Let the longitude and pericentre of the inner orbit satisfy $g_{1}  + h_{1}  = \pi$ without losing generality. This condition, combined with the nodal difference between the two orbits in the invariable plane reference system meets $h_{1}-h_{2}= \pi$, leads to some orbital elements of the inner binary can be eliminated from the general quadrupole Hamiltonian \citep{ford2000secular,naoz2013secular}.  Omitting constant terms, a `` simplified Hamiltonian" is obtained as follows,

\begin{equation}
H = (1 - e_1^2 + 5e_1^2{\sin ^2}{\Omega _2}){\sin ^2}{i_2},
\label{eq:sH}
\end{equation}

where, $e_1$ is the eccentricity of the inner binary orbit, and $\Omega _2$ is the longitude of outer orbit ascending node. $i_2$ is the inclination of the outer orbit relative to the orbital plane of the binary.

The range of $H$ is $\left[ {0,1 + 4e_1^2} \right]$.
Separatrix between circulation and libration of $\Omega _2$ is decided by the following,
\begin{equation}
H_c= 1 - e_1^2.
\label{eq:Hc}
\end{equation}

The ascending node of the test particle librates near ${90^ \circ }$ or $ - {90^ \circ }$ when $1 + 4e_1^2> H > {H_c}$. Simultaneously, the inclination oscillates.
The maximum and minimum of inclination appear when $\left| {{\Omega _2}} \right| = {90^ \circ }$.

A possible polar-ring debris disc was detected in 99 Herculis by the Herschel Key Program DEBRIS (Dust Emission via a Bias-free Reconnaissance in the Infrared/Submillimeter). This program characterized extrasolar analogs to the asteroid and Kuiper belts of the Solar System, which are collectively called  ``debris discs" \citep{pilbratt2010herschel}.
99 Herculis is a binary system, comprising an F7V primary (0.94 ${M_ \odot }$) orbited by a K4V secondary (0.46 ${M_ \odot }$) with the age of approximately 9.37Gyr. The semi-major axis of the binary is approximately 16.5 AU and its eccentricity is approximately 0.766. The $\textit{Herschel}$ Key Program DEBRIS discovered a debris disc located at 120 AU surrounding the binary. This disc is possibly a polar-ring debris disc, moving in a plane perpendicular to the binary pericenter direction and having a matchable lifetime with the star \citep{kennedy201299}.

\begin{figure}
	\includegraphics[width=\columnwidth]{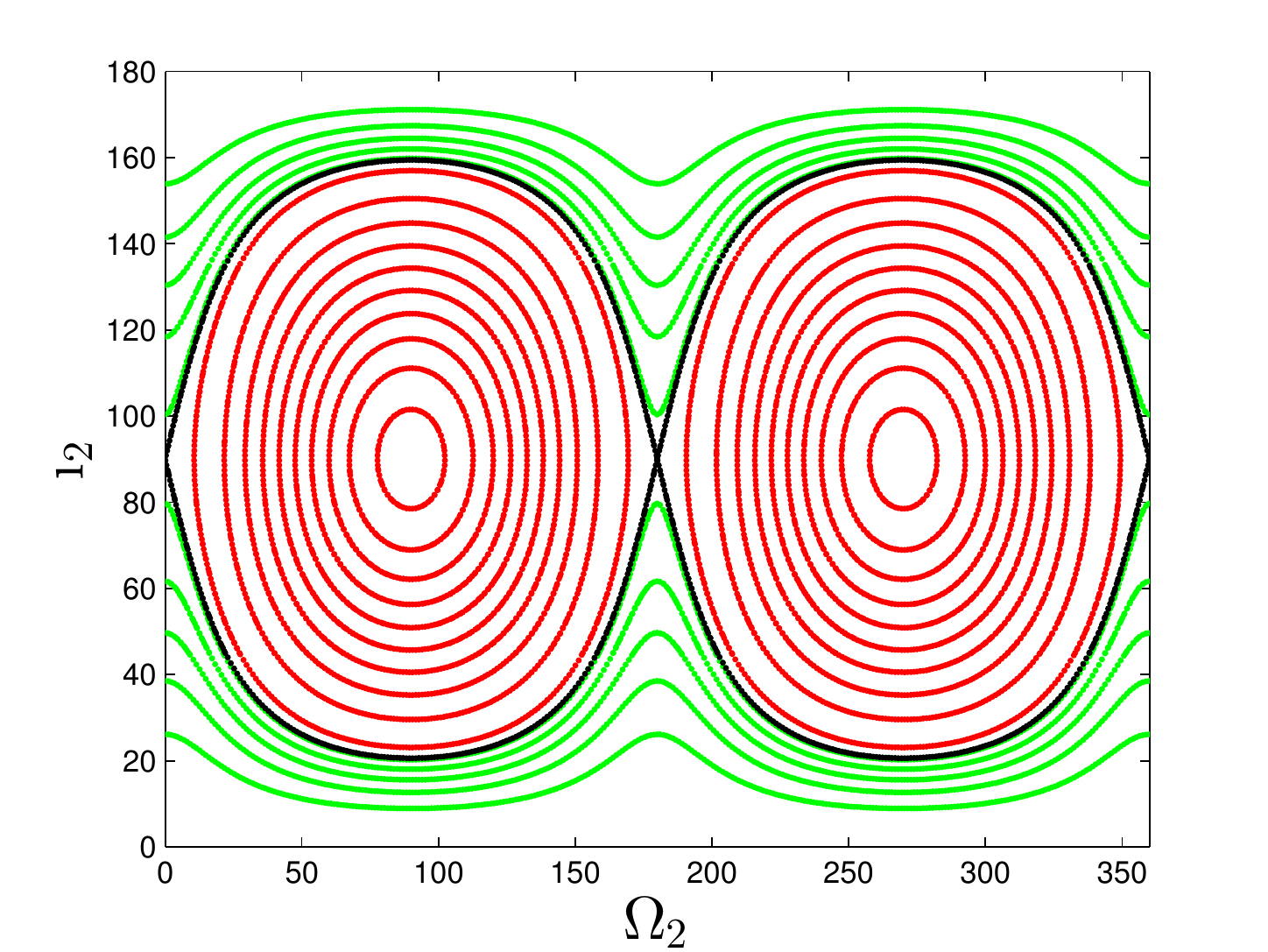}
    \caption{The phase space $(\Omega_2, i_2)$ of the simplified Hamiltonian
in equation~\ref{eq:sH} for 99 Herculis. The red region is the libration zone where the inclination and longitude of the outer orbit ascending node librate in a certain range. The green region is the circulation zone where the longitude of the outer orbit ascending node can circulate in $\left[0^\circ,360^\circ\right]$. The black line is the separatrix between circulation and libration in equation~\ref{eq:Hc}.}
    \label{fig:phase}
\end{figure}

The phase space $(\Omega_2, i_2)$ of the simplified Hamiltonian by equation~\ref{eq:sH} is plotted for $e_1 = 0.799$ (the binary eccentricity for 99 Herculis \cite{kennedy201299}) in Figure \ref{fig:phase}. When $\left| {{\Omega _2}} \right| = {90^ \circ }$, if the inclination of the orbital plane is larger than approximately $21^ \circ$, then the test particle will be deep in the libration zone with the inclination librating within $\left[21^ \circ,159^ \circ\right]$.

The theoretical analysis of the outer-restricted hierarchical problem is provided in this section to describe the circumbinary polar orbit in quadrupole approximation briefly. However, the long-term stability of polar orbits remains unknown. Only the quadrupole term is considered for the theoretical analysis based on the expansions of complete Hamiltonian in the ratio of the semi-major axis of the inner orbit to that of the outer one. The orbits may show different movements in the octupole Hamiltonian. For example, the eccentricity of near-polar orbits may be significantly excited \citep{li2014analytical}. Therefore, numeric calculations are necessary to investigate the stability of polar orbits with different orbital elements. Next, the stable region of test particles in polar orbits of the outer-restricted hierarchical three-body problem will be presented.

\section{The innermost stable orbits of test particles in polar orbits}
\label{sec:timescale}
\cite{pilat2003stability} recorded the escape time of inclined P-type orbits ($ 0^\circ \le i \le 50^\circ$) in binary systems with equal mass and eccentricity $0 \le e_1 \le 0.5$. If the binary eccentricity is larger than 0.35, then the critical inclination for libration will be smaller than $50^\circ$ for the equal-mass binary systems. Therefore, the simulations with $0.4 \le e_1 \le 0.5$ in the article \citep{pilat2003stability} may include the circulation and libration orbits. Their results revealed the absence of stability advantage of the orbits with high inclinations over those with low inclinations. However, the numerical simulations of the equal-mass binary with $0.4 \le e_1 \le 0.5$ in the paper of \cite{doolin2011dynamics} show that the high inclination of prograde orbits can be beneficial to the stability compared with the near coplanar prograde orbits. The results of \cite{pilat2003stability} and \cite{doolin2011dynamics} seem in conflict, but they are not. The critical inclination for libration, $i_c = \arcsin \sqrt {\frac{H}{{1 + 4e_1^2}}}$, is determined at $ {{h _2}} = {90^ \circ }$. \cite{pilat2003stability} did not fix the longitude of ascending nodes of planetary orbits specifically according to the separatrix between circulation and libration. They chose the ${\Omega _2}$ arbitrarily, the critical inclination turning into the libration zone can be quite larger than $i_c$, as revealed in Figure \ref{fig:phase}. The critical inclination turning into libration for the arbitrary node can be obtained by the following,

\begin{equation}
i_\Omega = \arcsin \sqrt {\frac{{{H_c}}}{{1 - e_1^2 + 5e_1^2{{\sin }^2}{h _2}}}}.
\end{equation}

\cite{doolin2011dynamics} presented a density plot of the stability measure across entire parameter spaces with the longest integration time ${5\times10^4}{T_1}$. Firstly, the peninsulas of instability in the libration region which appear symmetrically on either side of $i = \pi/2$ for the binary eccentricity $e_{1} \ge 0.3$ converge upon each other as $e_{1} \to 0.6$. This result implies that high binary eccentricity $e_{1} \ge 0.6$ could be more favourable to the stability of orbits in the libration region compared with $ 0.3 \le e_{1} < 0.6$. The innermost semi-major axis of stable orbits in the libration region is smaller than that of stable prograde orbits in the circulation region. This finding means that the critical semi-major axis of stable planets in polar orbits is smaller than that of coplanar prograde orbits, thus, raising the possibility of planet formation in the polar planetary discs. An important point to note is that the innermost stable orbits in the libration region appear when the inclination of orbit in the vicinity of ${90^ \circ }$ according to the stability maps in \citet{doolin2011dynamics} and \citet{cuello2019planet}. Then, we will find out the innermost stable orbits for different binary parameters, the mass ratio of binary star $\lambda$ and the binary eccentricity $e_1$.

In order to obtain the innermost stable polar orbits, a large number of numerical simulations need to be carried out. Without loss of generality, the semi-major axis of the binary orbit is set as 1 AU, and the binary eccentricity $e_1$ in the range $\left[ {0,0.8} \right]$ with scanned interval $\Delta e_1 = 0.05$. Simultaneously, the mass ratio of binary star $\lambda = \left[ {0.1,1} \right]$, where ${m_0} = 1{M_ \odot }$ and ${m_1}$ varied with an interval $0.05{M_ \odot }$. The semi-major axes of test particles range from 2 AU to 6 AU, with an interval of 0.01 AU. For a specific semi-major axis, eight inclinations were chosen from $\left[ {\arcsin \sqrt {\frac{H}{{1 + 4e_1^2}}} ,{{90}^ \circ }} \right]$ with equal interval. Eight mean anomalies distributed evenly in the range $\left[ {{0^ \circ },{{360}^ \circ }} \right]$ for each inclination. The initial longitude of ascending nodes $h_2$ was set as ${90^ \circ }$ in all cases. The orbits are circular initially, and the arguments of the pericenter are chosen arbitrarily. After ${10^7}{T_1}$, the initial semi-major axes of the innermost orbits that remained stable are regarded as the critical stable boundary.

\begin{figure}
	\includegraphics[width=\columnwidth]{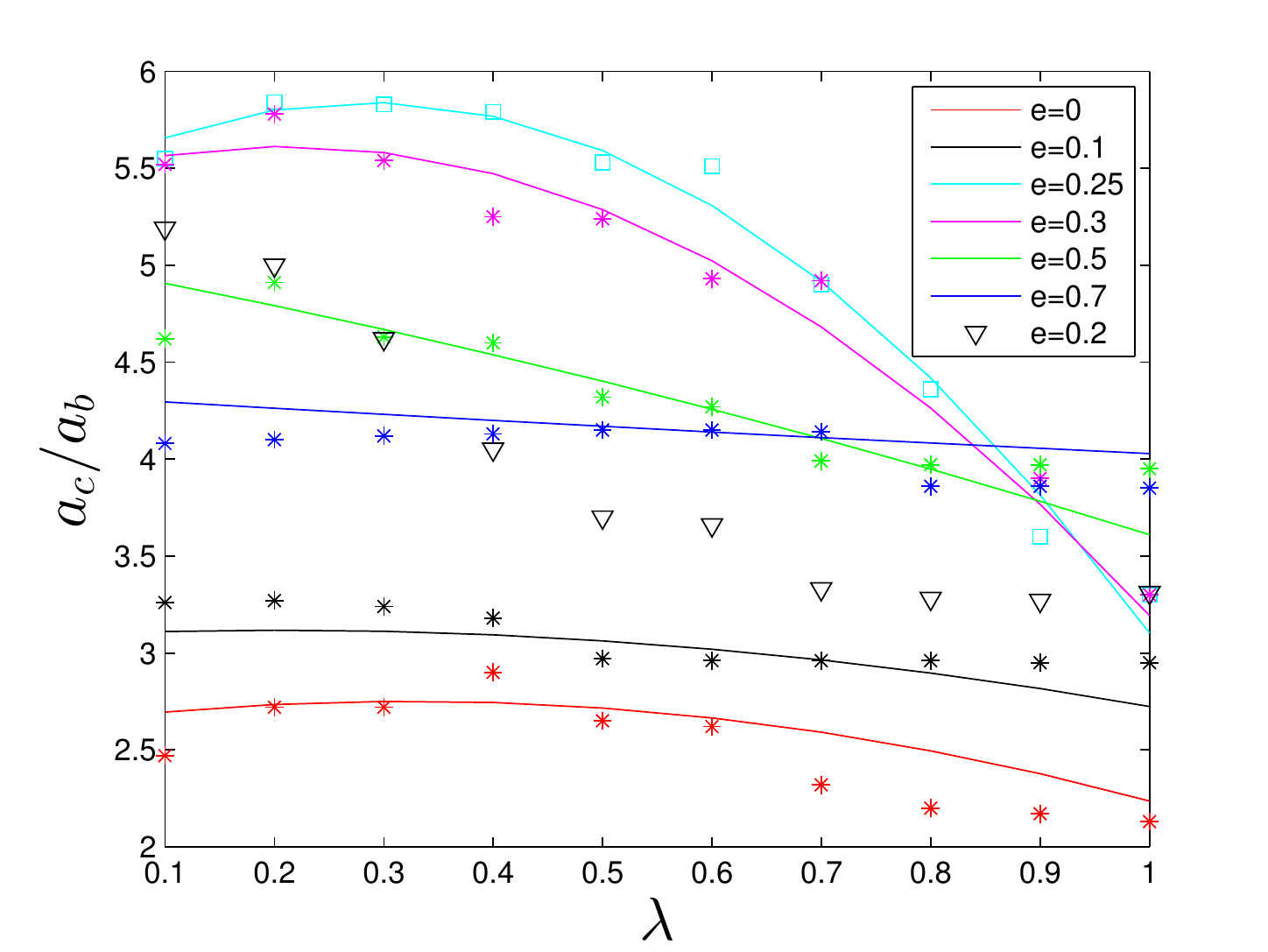}
    \caption{Several fitted curves of Equation \ref{eq:st1} and \ref{eq:st2}, as well as the respective raw data of the minimum stable radii of test particles in polar orbits around binaries with different eccentricities and mass ratios within ${10^7}{T_1}$.}
    \label{fig:st}
\end{figure}

The following empirical formulas are computed using multivariable linear regression analysis of the minimum stable semi-major axis $a_{\rm c}$ of test particles within ${10^7}{T_1}$ with different binary mass ratio $\lambda$ and eccentricity, 
\begin{equation}
\begin{array}{l}
\frac{a_c}{a_1} = 2.6338 + 4.9422{e_1} + 0.7237\lambda  - 3.6013{{e_1}^2} - 1.1215{\lambda ^2} \\
 - 5.2632{e_1}\lambda + 5.3516{{e_1}^2}\lambda +5.5769{e_1}{\lambda ^2} -5.6002{{e_1}^2}{\lambda ^2},

\end{array}
\label{eq:st1}
\end{equation}
for $0.1 \le \lambda \le 1,0.0 \le {e_1} \le 0.15$ and $0.65 \le {e_1} \le 0.8$ with the coefficient of determination $R^2 \approx 1.0$.  
\begin{equation}
\begin{array}{l}
\frac{a_c}{a_1} = 4.4096 + 6.7118{e_1} + 14.4955\lambda  - 10.9044{{e_1}^2} - 16.1753{\lambda ^2} \\
- 60.3628{e_1}\lambda  + 58.2231{{e_1}^2}\lambda +54.8823{e_1}{\lambda ^2} -46.2753{{e_1}^2}{\lambda ^2},
 
\end{array}
\label{eq:st2}
\end{equation}
for $0.1 \le \lambda \le 1,0.25 \le {e_1} \le 0.6$ with the coefficient of determination $R^2 \approx 0.8664$. Several fitted curves were plotted, and the corresponding raw data are presented in Figure \ref{fig:st}.

Apparently, the tendencies of minimum stable orbits with $\lambda$ are different for several eccentricity intervals in Figure \ref{fig:st}. The values of minimum stable semi-major axes of test particles moving in the polar orbits around the binary with ${e_1}=0.2$ are directly plotted in Figure \ref{fig:st} directly. Because the results with ${e_1}=0.2$ additional to the fitting processes of Equation \ref{eq:st1} and \ref{eq:st2} did not yield good fitting effects, which can be seen in Table \ref{tab:fitting}. The determination coefficient, F statistic, P value and error variance after multiple regression fitting for different cuts are shown in the table below. It seems more appropriate to divide the data simulation into two parts, one part with an eccentricity range of [0,0.15] and [0.65,0.8], and the other part with [0.25,0.6].

\begin{table*}
	\centering
	\caption{The parameters of multiple regression fitting for different cuts of binary eccentricities. }
	\label{tab:fitting}
	\begin{tabular}{|c|c|c|c|c|}
\hline
 $ e_1$     & $R^2$          & F statistic     & P value          & error variance \\
\hline
[0,0.8] &  0.72      &   547.13     &     0  & 0.20  \\
\hline
[0,0.15];[0.65,0.8] &  0.97      &   2917.30    &     0  & 0.015    \\
\hline
[0.25,0.6] &  0.90     &   879.56   &     0   &  0.052   \\
\hline
[0.2,0.6] &  0.76      &   350.71    &    3.37e-269   &  0.13   \\
\hline
[0.3,0.6] &  0.87      &   560.09   &     4.34e-296   &  0.057   \\
\hline
\end{tabular}
\end{table*}

For fixed binary eccentricity, the binary with comparable masses, $\lambda=1$, is favourable to the stability of polar orbits. These orbits around the binary with mild eccentricity ${e_1} <0.2$ can stably exist in the location closer to the binary compared with those around moderate and highly eccentric binaries. The stable boundary of polar orbits revolving the binary with high eccentricity${e_1} >0.6$ change slightly for different binary mass ratios. The numerical simulation on the stability of polar orbits revealed that finding planets in polar orbits at a close location in the binary systems with comparable masses and mild eccentricities ${e_1} <0.2$ is more hopeful.

\section{Accretion efficiencies of planetesimals in the polar-ring disc of 99 Herculis}
\label{sec:formation}
Although the polar-ring disc in 99 Herculis resolved in \citet{kennedy201299} was a debris disc that possibly underwent the stage of planet-formation. Another possible polar-ring disc in the young HD 98800 system comprises a gas-rich ``planet-forming'' disc\citep{ribas2018long, Kennedy_2019}.  \citet{zanazzi2018inclination} found that the timescale of the inclination evolution of circumbinary disc under the influence of disc warp profile and dissipative torque is shorter than the disc lifetime for typical disc parameters. Their finding implies that discs and planets may exist with high inclinations relative to the orbital plane of eccentric binaries. Based on hydrodynamic simulations, a series of articles worked by Martin and Lubow \citep{martin2017polar,martin2018polar,martin2019polar} pointed out that damped oscillations of tilt angle and longitude of ascending node of misaligned low-mass protoplanetary disc around an eccentric binary lead to a stationary state where the disc lies perpendicular to the binary orbital plane. They revealed how the evolution of a disc depends upon the parameters of the disc (mass, viscosity, temperature, and size) as well as the parameters of binary (binary mass ratio, orbital eccentricity, and inclination). The alignment time scale of the outer parts of sufficiently large discs may be longer than their dissipation time scale.

\subsection{Initial values of simulations}
The formation of planets in a polar planetary disc (namely polar case) is then examined after determining the stable boundary of test particles in polar orbits. 99 Herculis was selected as the numerical model based on the assumption of an extended and young polar planetary disc. The accretion efficiency of collisions among 30000 planetesimals with randomly chosen physical radii from 1-10 km and density of $3g \cdot c{m^{ - 3}}$ was investigated. Their orbital semi-major axes are evenly distributed in the range of 65-130 AU. There, $65 AU \approx  3.94 a_{1}$) is the numerical stable boundary which is consistent with 4.19 $a_{1}$ obtained by Equation \ref{eq:st1}. 

In two other cases, the coplanar disc of the same binary stars (namely coplanar case) and the general disc around a single star (namely standard case) were simulated for comparison. ``General disc" is based on the model of  ``minimum mass Solar Nebula", but with different scaling factors of solid surface density and solid enhancement beyond the ice line ($f_d$ and $f_ice$). The host in the standard case is a single star with a mass of 1.4 ${M_ \odot }$ (the total mass of 99 Herculis). 

Considering that the precession range of the orbital inclination in the polar disc is  $10^ \circ $ for the initial inclination distributed in the range ${85^ \circ } \sim {95^ \circ }$, in order to make the three planetary discs comparable, the orbital inclination ranges of the planetesimals in the coplanar disc and the standard disc were set   ${0^ \circ } \sim {5^ \circ }$ and initial longitudes of ascending nodes are chosen randomly. 

In the numerical simulation, the orbits of planetesimals in planetary disks need to be as close as possible to the state of natural dynamic evolution, so as to reflect the real collision results. If we start from circular orbits initially, the planetesimals in polar disc need about $6 \times {10^4}yr$  (about ${10^3}{T_1}$, where ${T_1}$  is the binary period) to get a natural state from fixed initial longitudes of ascending node at ${90^ \circ }$, and need $9 \times {10^5}yr$ (about $1.5 \times {10^4}{T_1}$ ) in the coplanar circumbinary disc from initial circular orbits. Computing time costs are expensive for 30000 planetesimals. So, we make efforts to let planetesimals move initially naturally to the greatest extent on the bases of dynamical properties. 

In the polar protoplanetary disc, initial inclination $i$  is chosen in the range ${85^ \circ } \sim {95^ \circ }$ randomly. From the phase space and analytical theory\citet{li2014analytical}, the inclination oscillates in the following range 
\begin{equation}
\left[ {\arcsin \sqrt {\frac{H}{{1 + 4e_1^2}}} ,\pi  - \arcsin \sqrt {\frac{H}{{1 + 4e_1^2}}} } \right].
\end{equation}

 So, we can obtain the range of initial longitudes of ascending nodes of planetesimals $\left[ {{\Omega _c},\pi  - {\Omega _c}} \right]$  for initial inclination $i$  which is chosen in the range ${85^ \circ } \sim {95^ \circ }$  randomly, where
\begin{equation}
{\Omega _c} = \arcsin \sqrt {{{\left( {\frac{{\left( {1{\rm{ + }}4e_1^2} \right){{\sin }^2}{i_c}}}{{{{\sin }^2}i}} + e_1^2 - 1} \right)} \mathord{\left/
 {\vphantom {{\left( {\frac{{\left( {1{\rm{ + }}4e_1^2} \right){{\sin }^2}{i_c}}}{{{{\sin }^2}i}} + e_1^2 - 1} \right)} {5e_1^2}}} \right.
 \kern-\nulldelimiterspace} {5e_1^2}}} for {i_c} = {85^ \circ }.
\end{equation}

In eccentric coplanar binary disc, considering the secular perturbation of binary, the initial eccentricities of planetesimals are chosen in the range $[0,{e_{\rm pump}}]$  arbitrarily \citep{moriwaki2004planetesimal}, where 
\begin{equation}
{e_{\rm pump}} \simeq \frac{5}{2}(1 - \frac{{{m_1}}}{{{m_0} + {m_1}}})\frac{{{a_1}}}{a_2}{e_1}. 
\end{equation}
The eccentricities of planetesimals are chosen from $0 \sim 0.05$ randomly in the general disc around the single star.

The inflated radii of planetesimals were adopted as $dr = 2{r^{\frac{3}{2}}}{(GM)^{\frac{{ - 1}}{2}}}d\nu $ (where $ d\nu=1m \cdot {s^{ - 1}} $, $M$ is the total mass of the binary) to trace the close encounter between planetesimals easily \citep{xie2009planetesimal}, which can ensure the accuracy of colliding velocity $\sim 1m \cdot {s^{ - 1}}$.

The inflated radius is artificially set to control the collision speed accuracy. In pure Kepler motion of two-body problem, the velocity of the celestial body moving around the host satisfies $v = {\left( {\frac{{GM}}{r}} \right)^{{1 \mathord{\left/
 {\vphantom {1 2}} \right.
 \kern-\nulldelimiterspace} 2}}}$ . The differential of the formula can be obtained, $dr = 2{r^{\frac{3}{2}}}{(GM)^{\frac{{ - 1}}{2}}}d\nu $ , which means that the velocity difference between two particles in Kepler motion at adjacent distance $dr$ is $d\nu $ . Therefore, the error between the relative velocity recorded here and the relative velocity of collision can be guaranteed by limiting the distance, $dr$, between two adjacent planetesimals before the collision. This error is proportional to the distance, that is to say, the relative velocity recorded when the distance between planetesimals is smaller will be closer to the relative velocity at the time of collision. In our code, we judge whether the two planetesimals are within the critical distance, $dr$. On the one hand, the collision requires the two planetesimals to be in close contact. On the other hand, we can control the error range between the relative speed calculated within the critical distance and the realistic collision speed.

%
%

During our simulations, the gas drag need not be considered. Because we found gas damping has little effect on the orbits of the planetesimals and collision results in our studies. On the one hand, for the outer part of the disc around 99 Her which locates far away from the binary, the local gas density is too small to dampen the planetesimals. The gas damping force is proportional to the gas density and the relative velocity between the gas and the planetesimal. According to the general model of gas disc \citep{sano2000magnetorotational, umebayashi2013effects}, at midplane, the gas drag force at 66 au is $2.8 \times {10^{{\rm{ - }}5}}$ times than that at 1 au in ``minimum mass Solar Nebula".  On the other hand, for the inner part of the disc, the dynamics of planetesimals are strongly excited by the binary, leading to the damping effect of the gas disc working inefficiently.  \citet{rafikov2013building} found that the gas drag does not resolve the fragmentation barrier issue in \textit{Kepler} circumbinary systems. Because fast relative precession of planetesimal and binary orbits results in inefficient planetesimal apsidal alignment. \citet{rafikov2013planet} also demonstrated that gas drag regulates eccentricity behaviour only for bodies with radii less than 1 km, which is below the adopted planetesimal size (1-10 km) in our article.


\subsection{Outcomes of Collision}
\label{sec:pengzhuang}

During the realistic collisions among planetesimals, the perturbation of gravitational force between the two planetesimals entering into the range of close encounters will gradually increase which can't be ignored. In our numerical simulations, gravitational interactions among planetesimals are neglected to save computation costs. We can use the physical parameters and the states of motion between the colliding planetesimals during entering into a close encounter to obtain the outcomes after collision approximately. The outcomes of collisions depend on the target mass ${M_{\rm targ }}$, the projectile mass ${M_{\rm p }}$, the target radius ${R_{\rm targ }}$, the projectile radius ${R_{\rm p }}$,  impact velocity ${V_{\rm i}}$ as well as impact factor $b$, which are defined as $ b = \sin \theta $, $\theta $ is the acute angle between the line of impact velocity and the barycenter line. The impact velocity vector ${\vec V_{\rm i}}$ is the relative velocity between the target and the projectile. During our process to calculate the outcome of the collision between two bodies, we set the heavier one ${M_{\rm targ }}$, and the lighter one ${M_{\rm p }}$. Figure \ref{fig:ratio} summarized the regimes and the calculations on the mass of the largest remnant after each collision briefly. The detailed process and scaling laws used to calculate maps of collision outcomes can be found in \citet{leinhardt2011collisions,stewart2012collisions}. 

\begin{figure*}
	\includegraphics[width=\columnwidth]{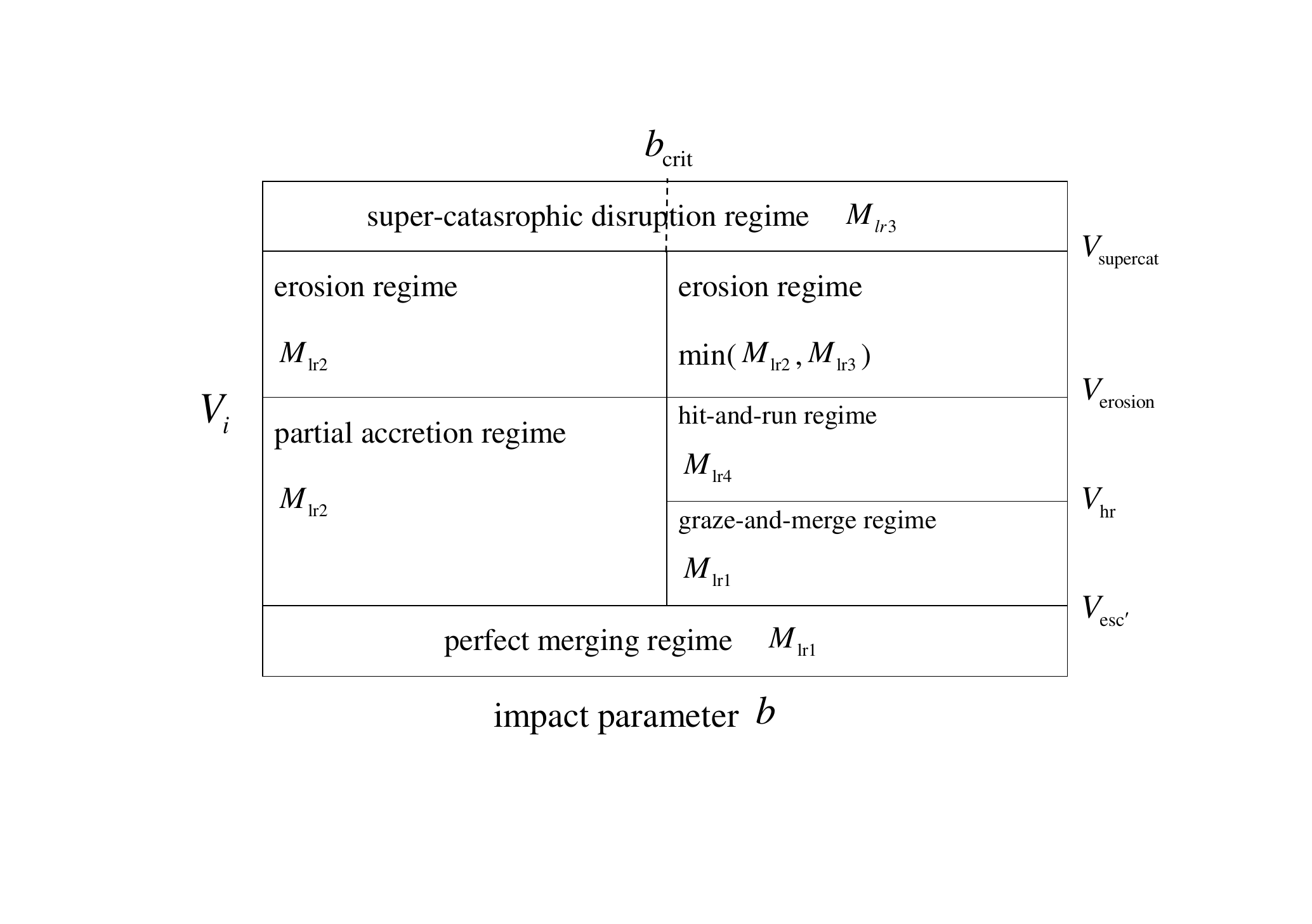}
    \caption{The outcomes of collisions mainly include seven regimes, which are determined by the impact velocity ${V_{\rm i}}$ as well as impact factor $b$. The corresponding mass of the largest remnant can be calculated by the  four equations below: $M_{\rm {lr1}}={M_{\rm p}} + {M_{\rm targ }},$
${M_{\rm lr2}} = \left( { - 0.5\left( {{{{Q_{\rm R}}} \mathord{\left/
 {\vphantom {{{Q_{\rm R}}} {Q_{\rm RD}^{'*} - 1}}} \right.
 \kern-\nulldelimiterspace} {Q_{\rm RD}^{'*} - 1}}} \right) + 0.5} \right)M_{\rm tot},$
${M_{\rm lr3}} = \frac{0.1}{1.8^\eta }{({Q_{\rm R}}/{{Q'}_{\rm RD}^*})^\eta}{M_{\rm tot}},$
${M_{\rm lr4}} ={M_{\rm targ }}.$ Other variates appearing in these equations and in this figure can be found in \citet{leinhardt2011collisions,stewart2012collisions}.}
    \label{fig:outcomes}
\end{figure*}

We have tried three methods to deal with the results of collisions. The first way, we assumed that the two planetesimals move in a straight line at a constant speed within the critical distance. A collision occurs when the distance between the planetesimals is less than the sum of the physical radii of the planetesimals. The statistical results show that almost no collisions have occurred, because this processing method ignores gravitational interactions between the planetesimals. 
The second way, all of the collisions are regarded as head-on collisions and the diversity of actual collision results is ignored. 
The third way considers the collision parameters and the calculation of the collision results is based on the coordinates and velocities of the planetesimals after they enter into the critical distance. This processing method will increase the proportion of hit-and-run collisions, which remains the two planetesimals undamaged. However, we can eliminate the artificially enlarged effect of the hit-and-run regime by sufficient collisions. Considering that the growth of planetesimals requires multiple collisions, we made stochastic simulations to find out the final mass of the largest remnant for a specific planetesimal after successive collisions. As long as the cumulative number of collisions is sufficient, the artificially enlarged effect of hit-and-run will be eliminated, and will not affect the final mass of the largest remnant. 
This is also the reason why we made statistical simulations in our article.

\subsection{Efficiency of accretion}

The efficiency of accretion $\xi$ is defined as follows,

\begin{equation}
\xi  = \frac{{{M_{\rm lr}} - {M_{\rm t }}}}{{{M_{\rm p}}}},
\label{eq:ea}
\end{equation}

where, $M_{\rm t}$ is the mass of the target, and $M_{\rm p}$ is the mass of projectile. $M_{\rm lr}$ is the largest mass of the remnant part after a collision between two planetesimals. The detailed calculations of  $M_{\rm lr}$ can be obtained in  Section \ref{sec:pengzhuang}. Clearly, $\xi=1$ means that the two planetesimals have merged together, thus demonstrating perfect accretion. $0<\xi<1$ means collisions lead to partial accretion, and $\xi<0$ implies target erosion. 

\begin{figure*}

	\includegraphics[width=\columnwidth]{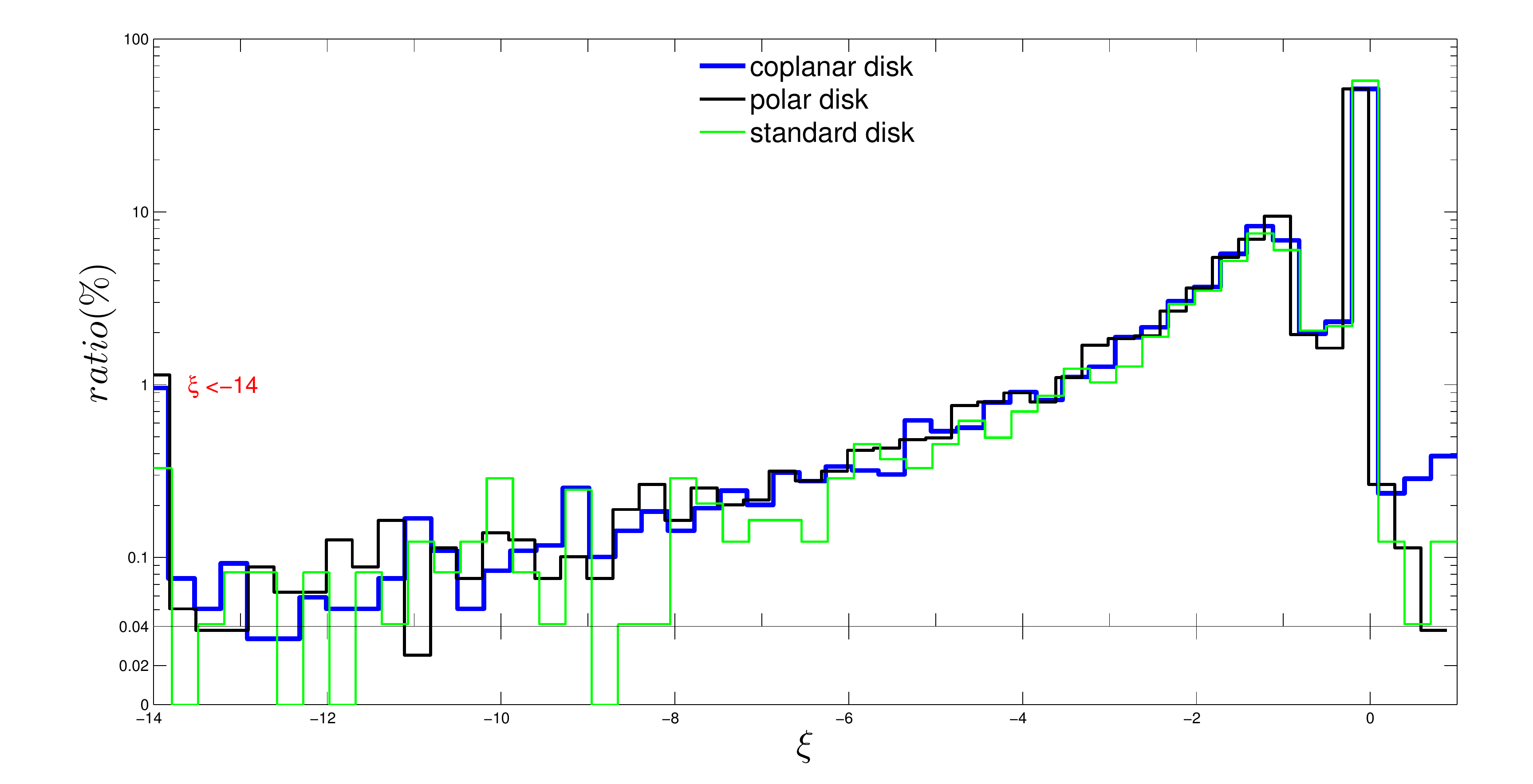}
    \caption{The collision outcomes among planetesimals in the polar protoplanetary disc of 99 Herculis system, the coplanar disc of the same binary stars (coplanar case), and the general disc around single stars with the mass of 99 Herculis system (standard case) for comparison. The coplanar case has the same initial conditions as the polar case except for inclination randomly chosen from ${0^ \circ }$ - ${5^ \circ }$. The standard case has the same initial conditions as the coplanar case except the host is a single star with mass 1.4 ${M_ \odot }$, the total mass of 99 Herculis.}
    \label{fig:ratio}
\end{figure*}

\begin{table*}
	\centering
	\caption{Brief look at the accretion efficiency  }
	\label{tab:tab1}
	\begin{tabular}{|c|c|c|c|c|c|}
\hline
                                & $\xi<0$ & $0<=\xi<0.5$ & $0.5=<\xi<1$ & $\xi=1$ & collision number\\
\hline
Polar case &  50.81\%      &   49.08\%     &     0.1011\%  & 0\% &7915 \\
\hline
Coplanar case &  49.15\%      &   50.29\%     &     0.4372\%  & 0.1261\%    & 11894\\
\hline
Standard case &  44.17\%      &   55.70\%    &     0.08233\%   &  0.04117\% &2429  \\
\hline

\end{tabular}
\end{table*}


The ratios of different efficiencies of accretion in the three discs are shown in Figure \ref{fig:ratio} and Table \ref{tab:tab1}. There is one thing worth noticing, the peaks appearing in the vicinity of $\xi=0$ include the overestimated part of hit-and-run. Hit-and-run can remain the planetesimals almost the same as before the colliding, and lead to $\xi \approx 0$.
The perfect accretion ($\xi =1$) in the polar-ring case is $0\% $, while that in the coplanar and standard case is approximately $0.1261\% $ and $0.04117\%$. Figure \ref{fig:ratio} shows that the accretion efficiencies among planetesimals that can cause accretion including part and perfect accretion in the polar protoplanetary disc are lower than either the coplanar binary disc or the single star disc. Meanwhile, the erosion collision ($\xi <0$) occurrence in polar-ring cases reveals slightly higher frequencies ($50.81\%$) than those of the two other cases ($49.15\%$ and $44.17\%$). Because of the expensive computer costs, we only simulated one group for each case, so there are no error bars. If enough numerical simulations are conducted, the efficiencies of accretion in the polar and coplanar circumbinary disc may be very close, presumably within error bars.
 In general, collisions among the planetesimals in the polar protoplanetary disc are not favourable for accretion compared with the single-star disc. 

According to the statistical data of collision results, we can basically deduce that there is a high probability that planetesimals cannot grow through collisions. In view of the following two reasons, we continued to carry out stochastic simulation: on the one hand, the proportion of hit-and-run results was overestimated in the process of collision result processing, and it was necessary to eliminate its influence through enough collisions; on the other hand, stochastic simulation can give out the maximum mass remained quantitatively after a series of collisions. 

Before stochastic simulations of successive collisions based on the ratios of efficiencies of accretion, collisional timescales require research, which determines the number of collisions that may occur within the age of 99 Herculis.

\subsection{Collisional timescale}

During the integration time ${10^{^5}}$yr, the numbers of collisions in the polar, coplanar and standard case are 7915,11894 and 2429 respectively as shown in Table \ref{tab:tab1}. Collisions among planetesimals occur most frequently in the coplanar case for the strong dynamical disturbing from the binary. The number of collisions in the polar case is moderate for mild gravitational disturbances from the binary in the perpendicular plane which affect the inclination more than eccentricity. The average numbers of collisions in the polar and coplanar circumbinary discs are respectively three and five times as many as those in the standard disc surrounding the single star.

In single-star systems such as our solar system, the objects of the Kuiper belt are distributed at $30$-$50$ AU from the Sun. The chance of collision among planetesimals will be remote considering the long period of objects and the shortage of disturbance. Generally, the collisional timescale of planetesimals with typical radius $R_p$ at semi-major axis $a$ in the standard disc around the single star with a mass $M_A$ can be obtained by
\begin{equation}
T_{\rm col}^S = \frac{2}{3} \times {10^4}f_{\rm d}^{ - 1}f_{\rm ice}^{ - 1}{\left( {\frac{\rm a}{{\rm AU}}} \right)^3}{\left( {\frac{{{M_A}}}{{{M_ \odot }}}} \right)^{{{ - 1} \mathord{\left/
 {\vphantom {{ - 1} 2}} \right.
 \kern-\nulldelimiterspace} 2}}}\left( {\frac{{{R_{\rm p}}}}{{\rm km}}} \right)yr,
\label{eq:ts}
\end{equation}
where, for ``minimum mass Solar Nebula", $f_{\rm d}=1$, and $f_{\rm ice}=4.2$ beyond the ice line, which lies in about 2.7 AU for solar-type stars.
According to \citet{xie2009bplanetesimal}, collisional timescales of planetesimals in coplanar circumbinary disc $T_{\rm col}^C$ and in polar circumbinary disc $T_{\rm col}^P$ can be estimated by
\begin{equation}
\begin{array}{l}
T_{\rm col}^C = T_{\rm col}^S \times {\left( {\frac{{n_{\rm imp}^C}}{{n_{\rm imp}^S}}} \right)^{ - 1}},\\
T_{\rm col}^P = T_{\rm col}^S \times {\left( {\frac{{n_{\rm imp}^P}}{{n_{\rm imp}^S}}} \right)^{ - 1}}.
\end{array}
\label{eq:ct}
\end{equation}
The impact rates $n_{\rm imp}^S$, $n_{\rm imp}^C$ and $n_{\rm imp}^P$ can be read out from our numerical simulations of three discs containing the same total numbers and physical radius distribution of planetesimals. 

\begin{figure}

	\includegraphics[width=\columnwidth]{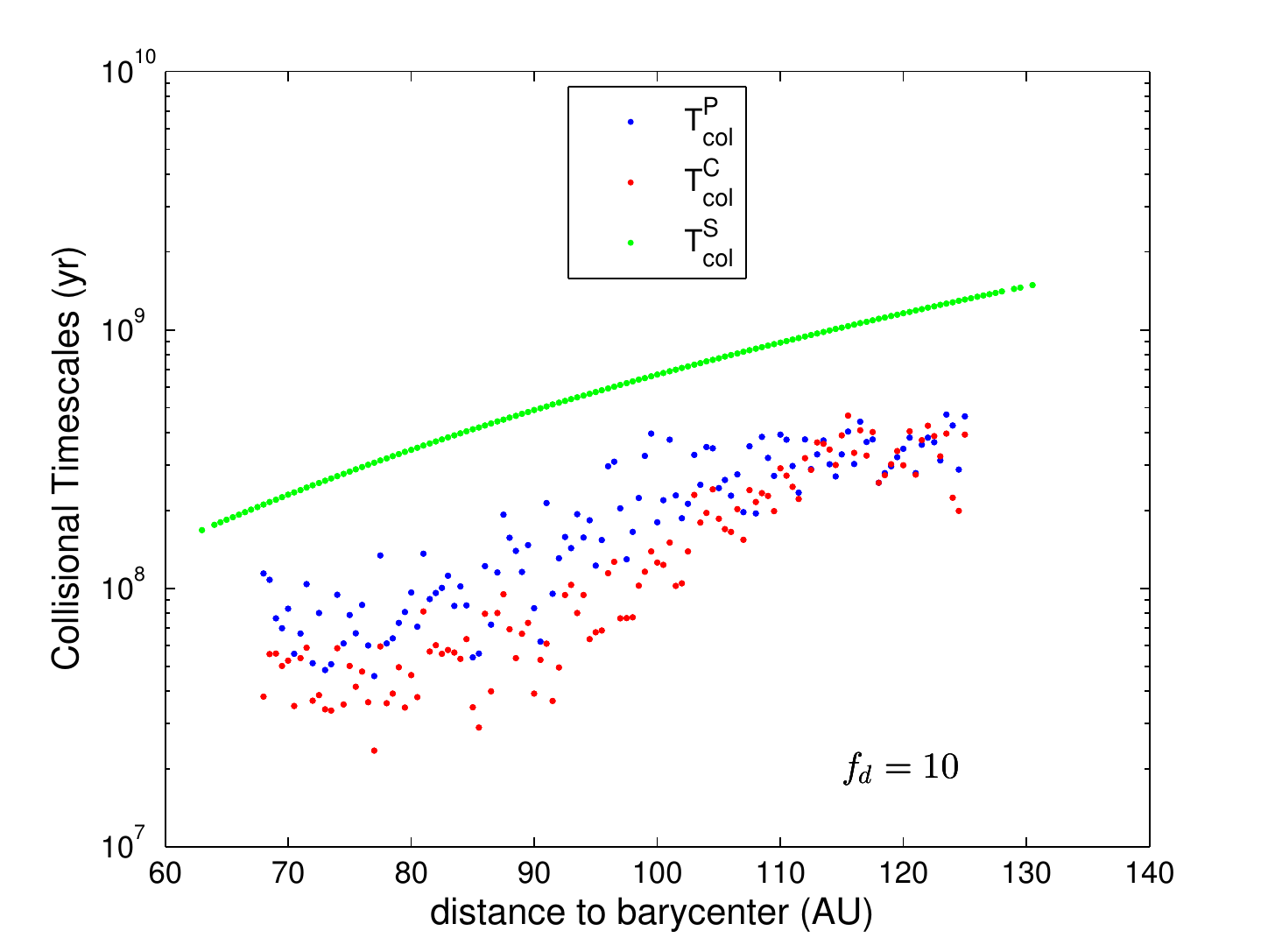}
    \caption{The collisional timescales of planetesimals in the polar circumbinary disc, $T_{\rm col}^{\rm P}$, and in the coplanar circumbinary disc, $T_{\rm col}^{\rm C}$ obtained by Equation \ref{eq:ct} combined with Equation \ref{eq:ts}. The scaling factor of solid surface density is set at a moderate value $f_d=10$.}
    \label{fig:ct}
\end{figure}
    
Although the impact rates of circumbinary planetesimals are several times than that in the standard disc, as shown in Figure \ref{fig:ct}, the collisional timescales $T_{\rm col}^{\rm P}$ of planetesimals in the polar circumbinary disc, and $T_{\rm col}^{\rm C}$ in the coplanar circumbinary disc with a moderate $f_d=10$ are longer than $10^7 yr$ which is beyond the observed dissipation timescale ($<6Myr$) of gas disc\citep{haisch2001disk,wyatt2008evolution}. \citet{ribas2015protoplanetary} confirmed that the dissipation timescale of the gas disc is directly related to the stellar mass. The gas disc around high-mass stars ($>2M_ \odot$) dissipated up to two times earlier than low-mass ones. That means, gas giants are hardly formed in such distant locations through core accretion for the collisional timescales are too long to absorb enough gas. Next, stochastic simulations were conducted to find out whether solid cores of protoplanets can form.

\subsection{Stochastic simulations}

However, \citet{batygin2016evidence} show that a possible distant giant planet with more than 10 Earth mass in the solar system can explain the cluster phenomenon of distant Kuiper Belt objects in the argument of perihelion and physical space.   
 \citet{Kenyon_2015} showed that super-Earth mass planets can form at $125$-$250$ au around solar-type stars from swarms of 1 cm-10 m planetesimals within 1-3 Gyr in the annuli with a mass of approximately $15{M_ \oplus }$ considering collisional damping.  However, this simulation did not include the gas drag, which is important for the vertical distribution and radial motion of dust and large bodies. The gas drag eventually becomes negligible once bodies of planetesimal size have formed. 

After obtaining the probability distribution law which has been shown in Figure \ref{fig:ratio}, we can obtain the probability density function. So that the efficiency of accretion of a random collision, $\xi$, can be obtained by the probability proportional to size sampling. The post-collision result can be calculated by the corresponding $\xi$. The mass of the largest remnant part $M_lr$ after this collision as one of the colliding planetesimals participates in the next collision. After N random samplings, the final mass of the largest remnant after N consecutive random collisions can be obtained.

The reasons for conducting stochastic simulation are as below. On the one hand, we can eliminate the overestimated hit-and-run part through enough collisions. On the other hand, the distribution of the final growth mass of planetesimals can be obtained quantitatively. How many collisions are enough? First of all, we can obtain the collision times $N_c$ by the collisional timescales within the age of 99 Herculis, about 9.37 Gyr.  We conduct sampling $100N_c$ in order to eliminate the overestimated hit-and-run part. The starting target planetesimal and the subsequent projectile ones have a physical radius of 10km.  There, we suppose there are sufficient planetesimals in protoplanetary discs. The detailed operations are as follows.

\begin{enumerate}
\item  The discrete distribution function of the accretion efficiency, $F(\xi ) = \sum\limits_{{\xi _n} < \xi } {{p_n}}$, can be obtained by the data of Figure \ref{fig:ratio}, for the distribution law as below.

\begin{tabular}{c|cccc}
  $\xi_{1}$ & $\xi_{2}$ & ... & $\xi_{n}$ &...\\
\hline
 $p_{1}$ & $p_{2}$ & ... & $p_{n}$ &... \\
\end{tabular}
 
\item  Generate a random number $\varepsilon$ from a uniform (0,1) distribution.

\item  For a $\varepsilon$, there must have an interval $(\xi_{n-1},\xi_{n})$, making $F({\xi _{n - 1}}) < \varepsilon  < F({\xi _n})$ with a probability $p_{n}$.

\item For a $\varepsilon$, an accretion efficiency $\xi={{\left( {{\xi _{n - 1}} + {\xi _n}} \right)} \mathord{\left/
 {\vphantom {{\left( {{\xi _{n - 1}} + {\xi _n}} \right)} 2}} \right.
 \kern-\nulldelimiterspace} 2}$ will be chosed. The mass of the largest remnant part after a collision will be calculated by Equation \ref{eq:ea}. The remnant continues to participate in the next collision.

\item If $\xi<-M_{p}/M_{t}$ leads to $M_{lr}<0$, which is a non-physical result, we need to regenerate a random number $\varepsilon$.

\item Sampling with probability proportional to size $100N_c$ times, then the final mass of planetesimal after enough collisions will be obtained.
\end{enumerate}

We carry out 10000 runs of probability proportional to size sampling for the collisions occurring in each kind of protoplanetary disc. The distribution of the final largest mass of the remnant part $M_{\rm lr}/M_{\rm ti}$ are listed in  Table \ref{tab:tab2}. $M_{\rm ti}$ is the initial mass of planetesimal before collisions. It is clear that the planetesimal can hardly grow bigger. In most cases, the final mass after $100N_c$ successive collisions is around the original mass at the beginning of collisions. It may be a possible mechanism to produce dust continuously, making a debris disc observed in 99 Herculis. An older debris disc ($>100Myr$) requires replenishment of dust through mutual collisions among a population of greater-than-kilometer-sized planetesimals. Some of the youngest(about 10-Myr-old) debris discs may be the remnant of protoplanetary discs\citep{wyatt2002collisional}. 99 Herculis is a main-sequence star with an age 6-10Gyr. So, the debris disc observed asks for a dust-generating process whether it is a long-lived debris disc surviving for Gyr timescales or a transient ring generated from a recent collision. Our stochastic simulations show that collisions of planetesimals in the polar protoplanetary disc of 99 Herculis will not make planetesimals grow, but produce dust steadily.

\begin{table}
	\centering
	\caption{The distributions of the final largest mass of the remnant part $M_{\rm lr}/M_{\rm ti}$ among $10000$ runs of stochastic simulations with $100N_c$ successive collisions. }
	\label{tab:tab2}
	\begin{tabular}{ccc}
\hline
                                       & $M_{\rm lr}/M_{\rm ti}$         & probability\\
\hline
                                       &$ <0.97$                  & $19.93\%$\\
Polar case                     &[0.97,1.05)   &$79.44\%$\\
                                       &[1.05,2]                 &$0.63\%$\\
\hline
                                               &$<0.97$                     &$13.74\%$  \\ 
Coplanar case                       &[0.97,1.05)          & $84.47\%$\\
                                              &[1.05,2]            &$1.79\%$\\
\hline
                                               &$<0.97$                  & $11.61\%$\\
Standard case                        &[0.97,1.05)      & $87.89\%$\\
                                              &[1.05,2]             & 0.5\%\\
\hline

\end{tabular}
\end{table}


Since the molecular gas is collocated with the dust in the debris disk, a new semi-analytical equivalent of the numerical model proposed by \citet{ kral2016self, kral2017predictions}, which assumes CO is produced from volatile-rich solid bodies. P-type exoplanets in circumbinary coplanar orbits have been detected by \textit{Kepler} \citep{martin2014planets}. A majority of these planets are located at the boundary of stable zones over long timescales \citep{holman1999long}. The stability limit is due to overlapping first- and second-order mean motion resonances with the binary, and is mainly influenced by the overlaps of three-body mean motion resonances in massive multi-planet systems \citep{wang2019influence}. However, the formation of planets in the circumbinary coplanar disc is possible for $a_{2} >20 a_{1}$ \citep{paardekooper2012not,marzari2012lunar,meschiari2012circumbinary}, or $a \approx 10-20 a_{1}$ considering the gravity of an axisymmetric disc which can strongly suppress the eccentricities of planetesimals beyond and facilitate the easy growth of planetesimals \citep{rafikov2013building}. Moreover, the critical radial distance beyond which planetesimal accretion is possible increases with the rising binary eccentricities \citep{scholl2007relative}. Thus, planet can hardly form around $65-130$ au ($3.94-7.88 a_{1}$) in the circumbinary coplanar protoplanetary disc with $a_{1} =16.5$ au and $e_{1} = 0.76$ of 99 Herculis. By comparison, the accretion efficiency of planetesimals in the circumbinary polar protoplanetary disc is lower than that in the coplanar one according to our simulation. So, we can infer the formation of planets in the inner region or around the location where the current debris disc exist is difficult in the system of 99 Herculis.

%

\section{Discussions and Conclusions}

\subsection{Discussions}

The innermost stable orbits simulated by \citet{cuello2019planet} and \citet{Chen2020MNRAS} locate much inner (about $2.5a_1$) for their integration time is $10^5T_1$ and $510^4T_1$, while $10^7T_1$ in our article. \citet{cuello2019planet} studied the evolution of misaligned circumbinary discs through hydrodynamical simulations. Viscous torques exerted by the binary make retrograde configurations easier to become polar than prograde circumbinary discs. 

\citet{childs2021formation} show that about five circumbinary planets form in polar and coplanar orbits in the vicinity of 5.4$a_1$. However, it simulates the late stage of the formation of planets with purely gravitational interaction for the Moon-sized planetesimals and Mars-sized embryos. That means, every impact can be regarded as a perfectly inelastic collision, which leads to perfect accretion. While the physical size of planetesimals in our simulations distribute in 1-10km. Various outcomes of collision including perfect accretion, partial accretion, hit-and-run, graze-and-merge and catastrophic disruption, super-catastrophic disruption, have been considered.

 A possible third component which is about 2.4 times as faint as 99 Herculis B was reported three times\citep{kennedy201299}. According to the three positions in sky coordinates, its possible stable orbits may be in Kozai cycles. The third component might be involved in the formation of the polar disk of 99 Herculis, which is interesting to further study. \citet{lepp2023polar} find out that polar circumtriple orbits only exist within a critical radius, outside which circulating orbits precess about the binary angular momentum vector. Through smoothed particle hydrodynamics simulations by \citet{Ceppi2023}, the wide range of disc inclinations in hierarchical systems with more than two stars may result from the secular oscillation of their orbital parameters. \citet{chen2022orbital} investigated the orbital dynamics of circumbinary planetary systems with two planets in polar orbits around the binary star. Under binary-planet and planet-planet gravitational interaction, the tilt angles of planets oscillate complicatedly.

The study on planetary formation in this paper is under the framework of the core accretion model. Some giant planets such as the four giants in HR 8799 system were detected far away from the central star through the method of imaging. If gravity instability gives rise to collapses in the solid component of the disc material, then giant planets can form in the outer regions of the protoplanetary disc. Planets formed beyond 100 au in solar-like gas discs through disc fragmentation can migrate inward, and produce giant protoplanets at a distance of a few tens of AU from the protostar via high-resolution numerical hydrodynamics simulations \citep{vorobyov2018gravitational}. 

Because typical collisional timescales in the polar disc are at least one order of magnitude higher than the dissipation timescale of the gas disk, gas giants are practically impossible to form by the core accretion model. In this model, the formation of planets mainly depends on the collisions among the planetesimals at the early stage. However, we couldn't exclude the possibilities of other models such as gravitational instability, pebble or dust-accretion producing planets or planet embryos.

\subsection{Conclusions}
The motion of planetesimals in polar orbits within the libration region of the outer-restricted three-body problem is simulated in this paper to study the stability of circumbinary polar orbits and the planetary formation in the circumbinary polar disc of 99 Herculis. Firstly, the empirical formulas of the stable critical semi-major axes of the polar orbits applied to $0.1 \le \lambda \le 1,0.0 \le {e_1} \le 0.15$ and $0.65 \le {e_1} \le 0.8$ in Equation \ref{eq:st1}, as well as $0.1 \le \lambda \le 1,0.3 \le {e_1} \le 0.6$ in Equation \ref{eq:st2}, are presented. Secondly, the collision outcomes, colliding timescales and stochastic simulation of successive collisions among the planetesimals in the polar protoplanetary disc of the 99 Herculis system are statistically analyzed. The statistical results show that the collisions of planetesimals (with the physical radius 1-10 km and density $3g \cdot c{m^{ - 3}}$) in the polar disc are the most unhelpful to the accretion and grow-up of planetesimals compared with the coplanar case and the standard case. Typical collisional timescales in the polar disc are at least one order of magnitude higher than the dissipation timescale of the gas disk. Furthermore, collisions of planetesimals in the polar protoplanetary disc of 99 Herculis will not make planetesimals grow, but produce dust steadily, which may explain the formation of the detected debris disc around 120 AU. Thirdly, considering the various outcomes of collisions among the planetesimals (1-10km), the performances of planetesimal growth by collisions in reference groups including the coplanar case and standard case are similar. The main differences between the three cases lie in the impact rate and the collision timescales.

\bibliography{Wang20230428}{}
\bibliographystyle{aasjournal}



\end{document}